\documentclass[english,twocolumn]{aastex63}
\usepackage[T1]{fontenc}
\usepackage[latin1]{inputenc}
\usepackage{subfigure}
\usepackage{amssymb}
\usepackage{float}
\usepackage{graphicx}
\usepackage{amssymb}

\makeatletter

\usepackage{natbib}
\usepackage{bm}
\usepackage{amssymb}
\usepackage{amsmath,amsfonts,amsthm,bm} 
\usepackage{babel}
\usepackage{graphicx}

\newcommand{\be}{\begin{equation}} 
\newcommand{\ee}{\end{equation}} 
\newcommand{\bea}{\begin{eqnarray}} 
\newcommand{\eea}{\end{eqnarray}}

\def\l{{\langle }}              %

\makeatother

\received{July 8, 2020}
\accepted{August 14, 2020}
\submitjournal{ApJL}

\begin{document}

\title{'Oumuamua as a Cometary Fractal Aggregate: the "Dust Bunny" Model}

\correspondingauthor{J. Luu}
\email{jane.luu@geo.uio.no}

\author{Jane X. Luu}
\affiliation{Centre for Earth Evolution and Dynamics, Department of Geosciences, University of Oslo}
\affiliation{Institute of Theoretical Astrophysics, University of Oslo \\
P.O. Box 1047, Blindern, NO-0316 Oslo, Norway}

\author{Eirik G. Flekk{\o}y}
\affiliation{PoreLab, the Njord Centre \\
Department of Physics, University of Oslo, P. O. Box 1048 \\
Blindern, N-0316, Oslo, Norway}

\author{Renaud Toussaint}
\affiliation{Institut de Physique du Globe de Strasbourg, University of Strasbourg}
\affiliation{PoreLab, the Njord Centre \\
Department of Physics, University of Oslo, P. O. Box 1048 \\
Blindern, N-0316, Oslo, Norway}


\begin{abstract}
The first known interstellar object, 1I/2017 U1 'Oumuamua, displayed such unusual properties that its origin remains a subject of much debate.
We propose that 'Oumuamua's properties could be explained as those of a fractal dust aggregate (a "dust bunny") formed in the inner coma of a fragmenting exo-Oort cloud comet.  Such fragments could serve as accretion sites by accumulating dust particles, resulting in the formation of a fractal aggregate.  The fractal aggregate eventually breaks off from the fragment due to hydrodynamic stress.  With their low density and tenuously bound orbits, most of these cometary fractal aggregates are then ejected into interstellar space by radiation pressure.
\end{abstract}

\keywords{Comets, comae, Oort cloud objects}

\section{Introduction}

Two  interstellar objects have visited the solar system in the last three years, 1I/'Oumuamua (\cite{mpec2017}) and 2I/Borisov (\cite{guzik2019}), and the two could not be more different.    Widely expected to be a comet, 'Oumuamua showed no sign of cometary activity  (\cite{jewitt2017,meech2017}), but its orbit nevertheless exhibited nongravitational acceleration (\cite{micheli18}).  In contrast, 2I/Borisov  is most definitely a comet, with cometary emission (\cite{opitom2019}), and optical colors similar to those of solar system comets  (\cite{jewittluu2019b, fitzsimmons2019, guzik2020}).  The stark difference between the two interstellar bodies suggests that whereas Borisov is probably a comet formed in a planetary disk, 'Oumuamua's origin lies elsewhere. Several models have been proposed to explain 'Oumuamua's origin, including tidal fragmentation  (\cite{cuk2018, zhang2020}), extinct fragment of a comet-like planetesimal  (\cite{raymond2018}), and an hydrogen-rich body formed in the dense core of a giant molecular cloud (\cite{seligman2020}).    
 \cite{flekkoy19}  showed that 'Oumuamua's unusual shape and ultra low density ($\rho = 10^{-2}$ kg/m$^3$, \cite{moromartin19b}) were consistent with a fractal aggregate with fractal dimension $D_f = 2.3-2.4$.   This fractal dimension is typical of an intermediate stage of planetesimal formation  (\cite{suyama08}), before the planetesimal becomes a compressed body with $D_f$ approaching the normal Euclidean dimension of 3.  It hints at an abbreviated accretion process; such a process could happen in the densest region of a cometary coma, near the nucleus's surface.  

In this work we build on the fractal framework for 'Oumuamua; we show how the object could have formed in the inner coma of an exo-Oort cloud comet, then escaped to interstellar space.  The paper is organized as follows.  Section 2 describes the conditions for accretion in the inner coma. The evolution of the fractal aggregate is described in Section 3, and in Section 4 we estimate the statistics for a population of cometary fractal aggregates.

\section{Aggregation in inner coma}

Consider a long-period comet nucleus with a radius $r_c = 1$ km, a gas-to-dust mass ratio of 1,  and an albedo of 0 for ease
of calculation (using a more typical albedo of 0.04 does not change our results). We will take the comet to 
be at heliocentric distance $R = 1$ AU,
although the process described here could potentially happen at any $R$ where water ice can sublimate.
We will also assume  $100\%$ active surface area, as is appropriate for long-period comets, and neglect thermal radiation.
At the subsolar point, the solar irradiance at a heliocentric distance $R$ [AU] is
$j_{sun} =\left( {1 AU}/{R} \right)^2  \mbox{1360 W/m$^2$ }$,
 so the mass loss rate per area is 
$Q = j_{sun} / {h_0} = 5 \times 10^{-4}$ kg m$^{-2}$ s$^{-1}$,
where   $h_0= 2.56 \times 10^6$ J/kg  is the latent heat of sublimation of water ice.
We assume the outflow to be isotropic, neglecting the difference
between the day and night side of the comet.  The energy balance equation
is then 
$h_0 {dM}/{dt} = \pi r_c^2 j_{sun}$, which yields the mass loss rate:
\be
{dM}/{dt}~{\mbox{[kg/s]}} = 1.7 \times10^3 \left( {1 \; AU }/{R} \right)^2 \left( {r_c}/ 1~{\mbox{km}} \right)^2.
\ee

We use the laws of conservation of mass, momentum and energy, and the Clausius-Clapeyron equation to calculate the gas density ,
gas velocity, gas temperature,  and gas thermal velocity (e. g., \cite{gombosi1985}) as a function of cometocentric
distance $r$ measured from the center of the nucleus.  
We assume that the gas dynamics completely determine the dynamics of the dust particles, which 
are lifted and accelerated by the gas flow. For large Knudsen numbers,  the drag force on a dust particle moving at
velocity $v_d$ depends both on the gas velocity $u$ and its thermal
velocity $v_{th} = \sqrt{k_B T/m}$ where $k_B = 1.38  \times 10^{-23}$ J/K is the Boltzmann constant, 
$T$ is the temperature, and $m$ is the molecule mass.
Assuming that gas drag is the main force affecting the dust particles, the dust equation of motion is then  
\be
m~d v_d/dt
= C_D  \pi r_0^2 \rho_g   v_{th} (u-v_d), 
\ee
\noindent where $C_D \sim 2$ is the drag coefficient (\cite{prialnik2004}), $\rho_g$ is the gas density, and $r_0$ the
particle radius. We solve this equation numerically, and plot the dust velocity $v_d(r)$ and  dust number density $n_d (r)$ in Figure 1a and
Figure 1b, respectively.   Figure 1a shows that, since the gas density decays quickly with distance, dust particles are soon decoupled from the gas as they travel away from the nucleus. 
Large particles have smaller terminal velocities because they accelerate more slowly, and do not have time 
to reach high velocities before they decouple from the gas.
Note that we have assumed that the dust loading had no effect on the gas, so that the gas and dust velocities calculated here are the maximum possible velocities. The rates of change of dust velocity shown in Figure 1a
imply accelerations $\sim 1 - 100$ m/s$^2$ for the different particle sizes, which are 4-6 orders of magnitude larger than the acceleration of gravity; for this reason gravitational forces are neglected in the rest of this work.

\begin{figure}[h!]
\plotone{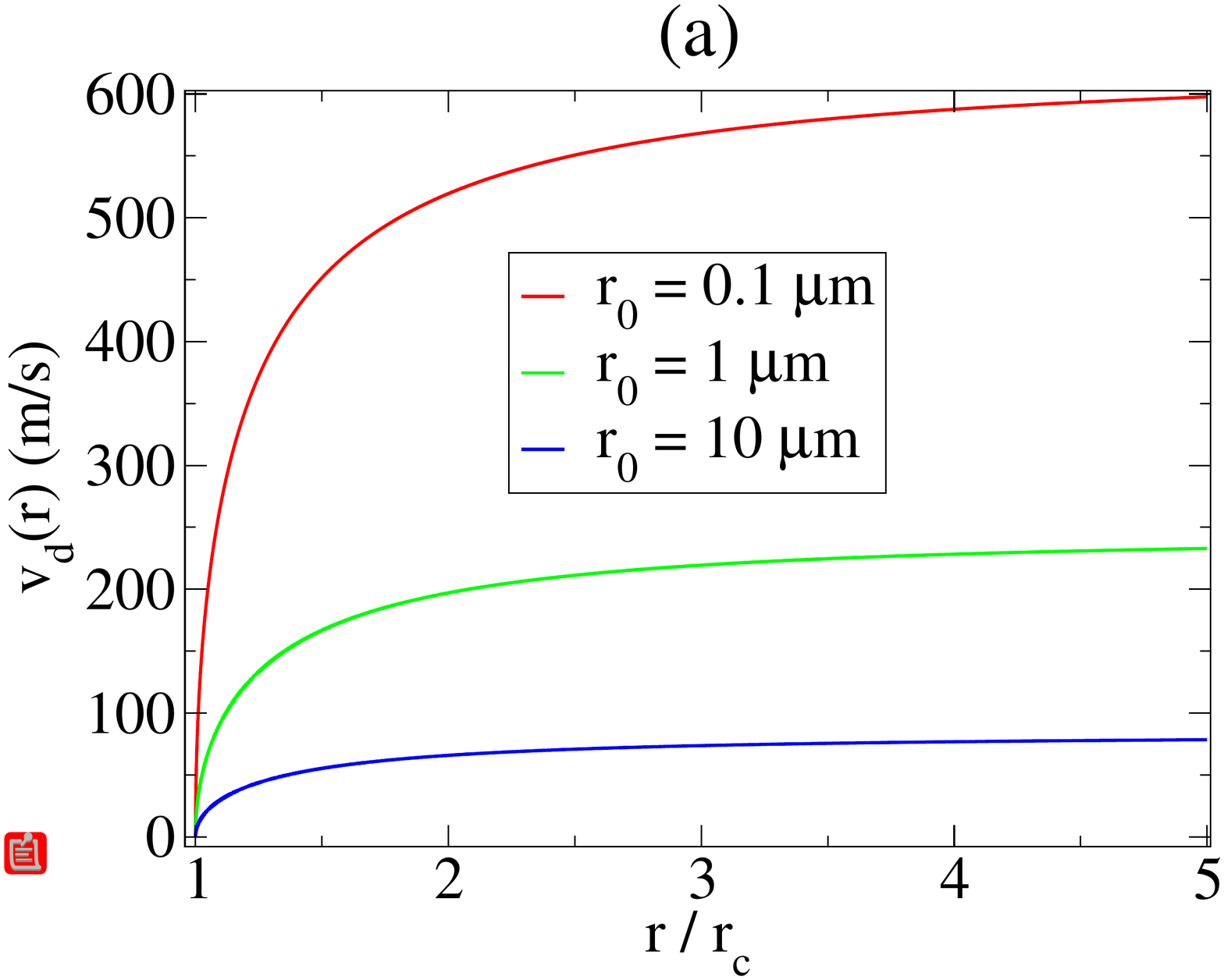}
\plotone{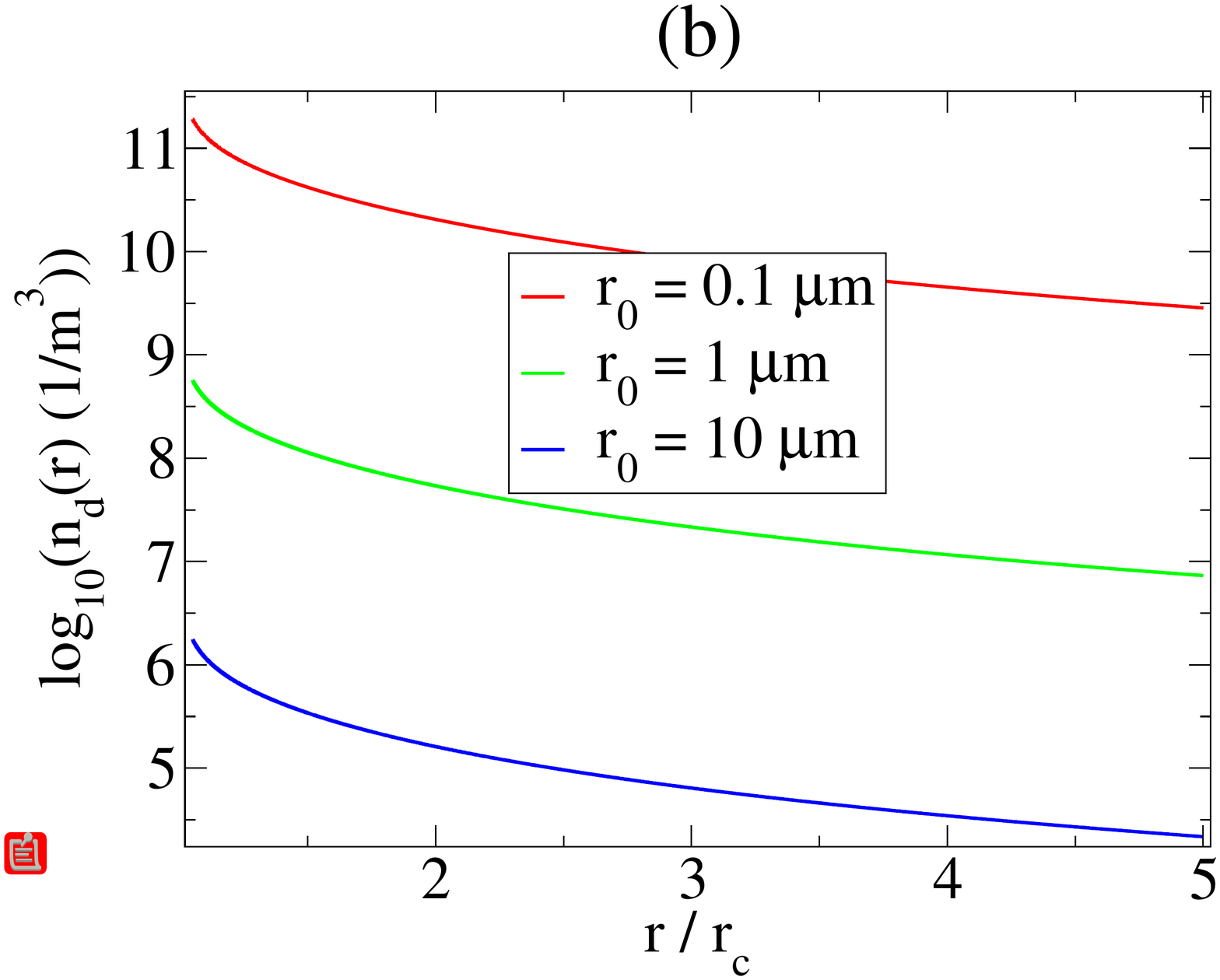}
\caption{Dust dynamics as function of cometocentric distance, at $R=$ 1 AU.  (a)
  Dust velocity for different $r_0$. (b) Dust number density for different $r_0$. \label{vd}}. 
\end{figure}

It is well known from  simulations of  granular gases
that colliding particles dissipate their relative velocities and form clusters, even in the absence of adhesion forces  (\cite{goldhirsch93}).
In the inner coma,
even though micron-size dust particles are traveling at $\sim 200$ m/s, the relative velocities between similar size particles may be small enough for aggregation to take place.  We note that both fluffy dust particles ($D_f \sim 1.8$)  and  
more compact particles ($D_f \sim 2$) have been observed in the coma of comet Churyumov-Gerasimenko,
suggesting particles that have undergone different levels of aggregation and collisional compacting  (\cite{mannel2016, mannel2019}).

The key parameter that determines whether aggregation can start is the
mean free path $\lambda$ between dust particles.  Assuming unit dust particles with radius $r_0$, the mean free path 
can be written as $\lambda (r) = {1}/[4\sqrt{2}{n_d (r)  \pi r_0^2}]$.
Hereafter we use  $r_0 = 1~\mu$m as the unit particle radius (\cite{mannel2019}), with
 mass $m_0 = (4/3) \pi \rho_0 r^3_0 = 4 \times 10^{-15}$ kg, where $\rho_0 = 1000$ kg/m$^3$.
The mean free path $\lambda (r)$ is plotted in Figure 2.   Clearly, when the mean free path is larger than the distance traveled by the dust particle, the particle will tend to escape without forming aggregates, so this may serve as the criterion for growth.  Assuming normal ejection for the dust particles, the growth criterion can be written as $ \lambda (r) < (r -r_c) $,  where  $r - r_c $ is the distance traveled by a dust particle.  The line $ \lambda (r) = (r -r_c) $ is plotted as a dashed line in the Figure; aggregation 
will only occur for unit particles that lie below this line. The Figure shows that, between particles with radius $r_0 \le 1 \ \mu$m, collisions will occur out to $r \sim 2.5r_c$; as dust particles grow larger, collisions will involve larger particles.

\begin{figure}[h!]
\plotone{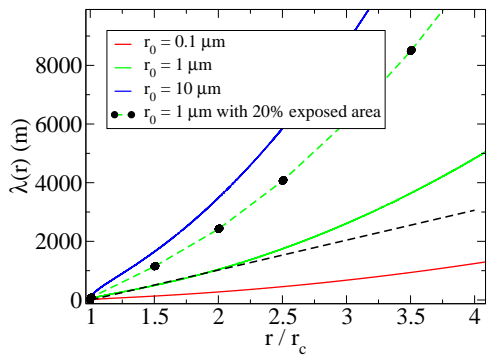}
\caption{Mean free path at $R=$ 1 AU for different $r_0$, as 
a function of cometocentric distance $r$ normalized by nucleus radius $r_c$. The black dashed line is the $\lambda = (r -r_c)$
line. The green line with black dots is also for $r_0 = 1~\mu$m, but with a 
surface flux corresponding to a nucleus with 20\%  fractional active area. \label{mfp}}. 
\end{figure}

\section{Formation and evolution of fractal aggregate}

However, collisions alone will not produce very large aggregates, since the particles move so quickly through the
inner coma.  But this region is most likely not populated by dust particles alone.  Comets are prone to fragmentation, 
particularly long-period comets (e.g., comet Hyakutake (\cite{weaver01}), comet C/2019 J2 (\cite{jewittluu2019a}),
so we suspect that the inner coma must also contain many small, meter- to 10m-size, fragments that often go undetected. 
For example, the numerous meter-size boulders strewn on the surface of 
comet Churyumov-Gerasimenko (\cite{cambianica2019}) may represent those fragments that failed to reach escape velocity. 
These fragments would accumulate dust particles as they travel through the coma, acting as accretion sites. Cometary dust is known to be fractal in nature (\cite{mannel2019}), so the accumulating particles are expected to result in a fractal body, hereafter referred to as a cometary fractal aggregate (CFA).  

To serve as an accretion site, a cometary fragment must be large enough to absorb the momentum of the impinging dust particles.  To calculate the minimum size of a fragment acting as accretion site, we assume the CFA to be identical to 'Oumuamua.  The latter's dimensions are uncertain, due to the unknown albedo.  Assuming an albedo of 0.04, \cite{jewitt2017} and \cite{meech2017}  reported roughly similar dimensions of 460 m $\times$ 70 m $\times$ 70 m, and 800 m $\times$ 80 m $\times$ 80 m, respectively.  A
larger albedo would yield correspondingly smaller dimensions.  For example, \cite{mashchenko2019} assumed an albedo of 0.1 and found that 'Oumuamua's lightcurve could be well fitted by either an oblate ellipsoid with dimensions 115 m $\times$ 111 m $\times$ 19 m, or a prolate ("cigar-like") 342 m $\times$ 42 m $\times$ 42 m ellipsoid.  The non-detection of the object by the Spitzer Space allowed for an object as large as 1 km in the largest dimension (\cite{trilling18}).   The consensus from optical measurements is that 'Oumuamua was several $\times 100$ m along its longest axis, with at least one large ($\sim 6:1$) axis ratio.  The uncertain dimensions are not important to the model we are presenting here, as we simply seek to show that an elongated fractal object several hundred meters in size could form in a cometary coma.  In this paper, we adopt the following parameters for the CFA: mass $M_{CFA} = 10^4$ kg, surface mass density 3 kg/m$^2$, and semimajor axes 230 m $\times$ 35 m $\times$ 35 m, with the long axis aligned with the gas flow direction, the short axis perpendicular to the flow.  

With a gas-to-dust mass ratio of 1, the dust flux near the nucleus's surface at $R = 1$ AU is $Q_d = dM/dt~/(\pi r_c^2) = 5 \times 10^{-4}$ kg m$^{-2}$ s$^{-1}$.
This mass flux rate impinging on a circular area of radius $b = 35$ m implies an accretion rate of 2.1 kg/s, thus requiring $10^4$ kg / (2.1 kg/s) = 4800 s for the CFA to grow to 'Oumuamua's size.  If the growth region is $\sim 1$ comet radius
 above the surface  (Figure 2), this implies a maximum fragment velocity $V_{frag}=$ 1 km/4800 s = 0.2 m/s.
For the fragment to absorb the dust momentum and not accelerate above $V_{frag}$, 
it must then have a minimum mass $M_{frag}= (v_d/V_{frag}) M_{CFA} =10^7$ kg, where $v_d = 200$
m/s (Figure 1a). This mass corresponds to a  diameter $\sim 13$ m if the fragment has density $\rho_0 = 10^3$ kg/m$^3$.  Comet fragments that have been studied in detail
showed remarkably similar composition, contrary to the expectation that they should show compositional 
variations if they come from different depths in the nucleus (\cite{dellorusso2007}).  One possible explanation for this compositional homogeneity is the accumulation of dust aggregates on their surface, as we are proposing here.

\subsection{Collisional deformation}

As the fragment moves through the coma at  $0.1 - 10$ m/s  (\cite{sekanina1978}), it will be bombarded
by single dust particles and small clusters, predominantly on the side facing the comet.  
These projectiles will arrive at $\sim 200$ m/s, and may break as they strike the fragment, but 
their abundance should still lead to growth by mass transfer (\cite{meisner2013, johansen2014}).
Experiments have shown that high speed collisions were an important mode of accretion,
allowing large bodies bombarded by smaller particles to grow further by scavenging the smaller aggregates.

Once small clusters are established on the cometary fragment, growth should proceed much more rapidly, resulting in a fractal body much like
a ballistic cluster-cluster aggregate (BCCA).  The main growth  mechanism is now no longer mass transfer, but simple
aggregation, due to the fact that BCCAs are porous, highly compressible bodies and 
therefore very effective absorbers of energy (\cite{donn1991}).   
Early on, the CFA may display filamentary structures typical of BCCAs, and is characterized by a fractal dimension $D_f =
1.8 - 2$ (\cite{suyama08}); however, as it grows from impacts, the CFA will compress and deform, 
resulting in a fractal dimension closer to $D_f = 2.5$ (\cite{suyama08}).   We show below that the CFA can survive the deformation given the impact energies involved.

We assume that the dust particles striking the CFA initially have fractal dimension $D_0 = 2$ (\cite{suyama08}).
In order to accommodate deformation, the CFA's internal strength 
 and volume must be able to withstand the impact energy.
The impact energy is equal to the projectile's kinetic energy, $E_p = M_p v_d^2/2$, where  $M_p = (4\pi /3) \rho_0 (r_p/r_0)^{D_0} r_0^3$ is the projectile's mass, and $r_p$ its radius.  The energy needed to deform the CFA is $E_{def} =  \overline{\sigma}_{yield}   V_{def}$, where
$\overline{\sigma}_{yield}$ is the average yield stress of the CFA,  $V_{def}=(4\pi /3)~l_{def}^3$ is the volume of the deformation, and $l_{def}$ the radius of a spherical deformation volume.  The average yield stress derives from the stress needed to break the bond between two $r_0$-size particles (\cite{johnson71}); for 'Oumuamua it was found to be $\overline{\sigma}_{yield} = 0.2$ Pa (\cite{flekkoy19}).  
Equating the two energies, 
\be 
M_p v_d^2/2 = \overline{\sigma}_{yield}   V_{def}, 
\ee

\noindent we can solve for $l_{def}$ in terms of the projectile's size, velocity, and $\overline{\sigma}_{yield}$.  With $D_0= 2$,
\be
l_{def} = \left[ \rho_0 v_d^2 r_0 r_p^2 /(2 \overline{\sigma}_{yield}) \right]^{1/3}.
\label{ldef1}
\ee

%

We are interested in the maximum deformation, so we need to calculate the maximum $r_p$, i.e., the size of the largest projectile  that could impact the aggregate.  The projectile grows according to $dM_p/dt = \pi r_p^2 Q_d$.
Since $M_p$ scales as $r_p^{D_0}$, we can write
$\pi r_p^2 = \pi  r_0 ^2 \left(M_p/m_0\right)$.  The growth equation can then be rewritten as
\be
dM_p/dt = M_p / \tau_p, 
\ee

\noindent where $\tau_p  = m_0 /(\pi r_0^2 Q_d) \sim 2.5$ s is the characteristic growth timescale for the projectile.

The solution of this equation is  
\be
M_p(t) = M_p(0) \exp \left(t/\tau_p \right), 
\ee
\noindent with the boundary condition $M_p(0) = m_0$.
The time available to a projectile to grow is roughly equal to 
the transit time through the accretion region, $t_0 = r_c/{v}_d \sim 5$ s when  $v_d =  200$ m/s.  Since $r_p \sim M_p^{1/2}$, we can write
\be
 r_p (t) = r_0~\exp \left[ t_0/(2 \tau_p) \right] = r_0~\exp \left[ r_c/(2 v_d \tau_p) \right].
 \ee
 
\noindent Inserting this expression into Equation \ref{ldef1}  yields
\be
l_{def} = r_0 \left[ \rho_0 v_d^2/(2\overline{\sigma}_{yield} ) \right]^{1/3} \exp \left[ r_c/(3v_d \tau_p) \right].  
\label{ldef2}
\ee

\noindent For $R$ = 1 AU, $l_{def} \sim 1$ mm and can easily be accommodated by the growing aggregate.  This is because the transit time through the accretion zone is so short that $r_p$ does not have time to grow very large ($\sim$ a few $\times~1~\mu$m).  However, the exponent in Equation \ref{ldef2} contains $\tau_p$, which depends on the heliocentric distance $R$, so that the deformation length grows exponentially with decreasing $R$.  For $R << 1$ AU,  $l_{def}$ becomes so large that the collisions become destructive and growth is not possible.

\subsection{Break-off from fragment}

Eventually, preferential growth along the flow direction should lead to an elongated, dangling CFA (Figure  3). 
A cometary fragment with such a dangling tail is an unstable configuration, 
akin to a feathered dart with a tail wind, and the hydrodynamic stress from the coma
environment will eventually cause the CFA to buckle and break off. 
This occurs when the hydrodynamic stress force exceeds the internal mechanical strength, and it sets an upper limit on the semi-minor axis $b$ (the dimension perpendicular to the gas flow), as we show below.

We consider the situation where the CFA is still growing and pointing upstream,
exposing an area $\pi b^2$ to the gas flow.   The dominating force that acts on the entire length of the aggregate is the hydrodynamic drag due to the gas, because the  gas velocity is so much larger than the dust velocity,
and thus carries more momentum.  Near the nucleus, the drag force is given by 

\be
F_D = ( {\pi b^2C_D}/{2})\rho_g  u^2,
\ee

\noindent where $\rho_g$ is the gas density and $u$ is the gas velocity.
This drag force is balanced by the aggregate's internal stress force 

\be
F_{int} = \overline{\sigma}_{yield}~\pi b^2.
\ee

\noindent $\overline{\sigma}_{yield}$ is the product of the single-link yield stress $\sigma_{yield}$ (\cite{johnson71}) and the solid fraction (filling factor) $\phi$:
\be
\overline{\sigma}_{yield} = {\sigma}_{yield} ~\phi, 
\ee

\noindent with $\phi = (b/r_0)^{D_f-3}$.  The internal stress force can then be written as 

\be
F_{int} = \pi  b^2 \sigma_{yield} \left( b/r_0 \right)^{D_f-3}.   
\ee

\noindent With  the CFA characterized by $D_f=2.3$, the internal strength $F_{int}$ scales as $b^{1.3}$ while the drag force $F_D$ scales as $b^2$. This  means that, as $b$ increases, the drag force will eventually overwhelm the
internal strength, and the structure will start to deform at some critical $b_{max}$.
This deformation may take the form of pure compression, but rodlike
structures under compression tend to buckle then break off.   
$b_{max}$ can be calculated by  equating the drag force with the internal force: 

\be
\left( \pi b^2 C_D/2 \right)~\rho_g  u^2
= \pi \sigma_{yield}~b^{1.3}~r_0^{0.7},  
\ee

\noindent which yields

\be
b_{max} =  r_0 \left[
   { 2\sigma_{yield}
  }/( \rho_g  u^2) 
\right]^{\frac{1}{0.7}}  = 50~\mbox{m}
\ee

\noindent  where we have assumed $\sigma_{yield} = \mbox{50 kPa} \left({\mbox{1 $\mu$m}}/ {r_0} \right)$ (\cite{flekkoy19}), $r_0 = 1~\mu$m, $C_D \sim 1$, and the gas parameters at $R = 1$ AU, $\rho_g = 6 \times 10^{-7}$ kg/m$^3$ and  $u= 825$ m/s.  

The close agreement between this predicted $b_{max}$ and the observed value of $b = 35$ m must be considered fortuitous,
given the uncertainties in the model parameters.  Nevertheless, the agreement between model and measurement lends support to 
the assumption that 'Oumuamua's dimensions are governed by hydrodynamics in the coma. 

In the discussion above we have assumed that the cometary fragment's rotation period was much longer than the 4800-sec accretion time of the CFA, allowing the CFA to develop an elongated shape.   If the opposite is true, the CFA would acquire a disk shape, instead of a cigar-like shape.  Both shapes are consistent with the observed lightcurve  (\cite{mashchenko2019}); for simplicity, we will stay with the cigar shape model.

\begin{figure}[h!]
\plotone{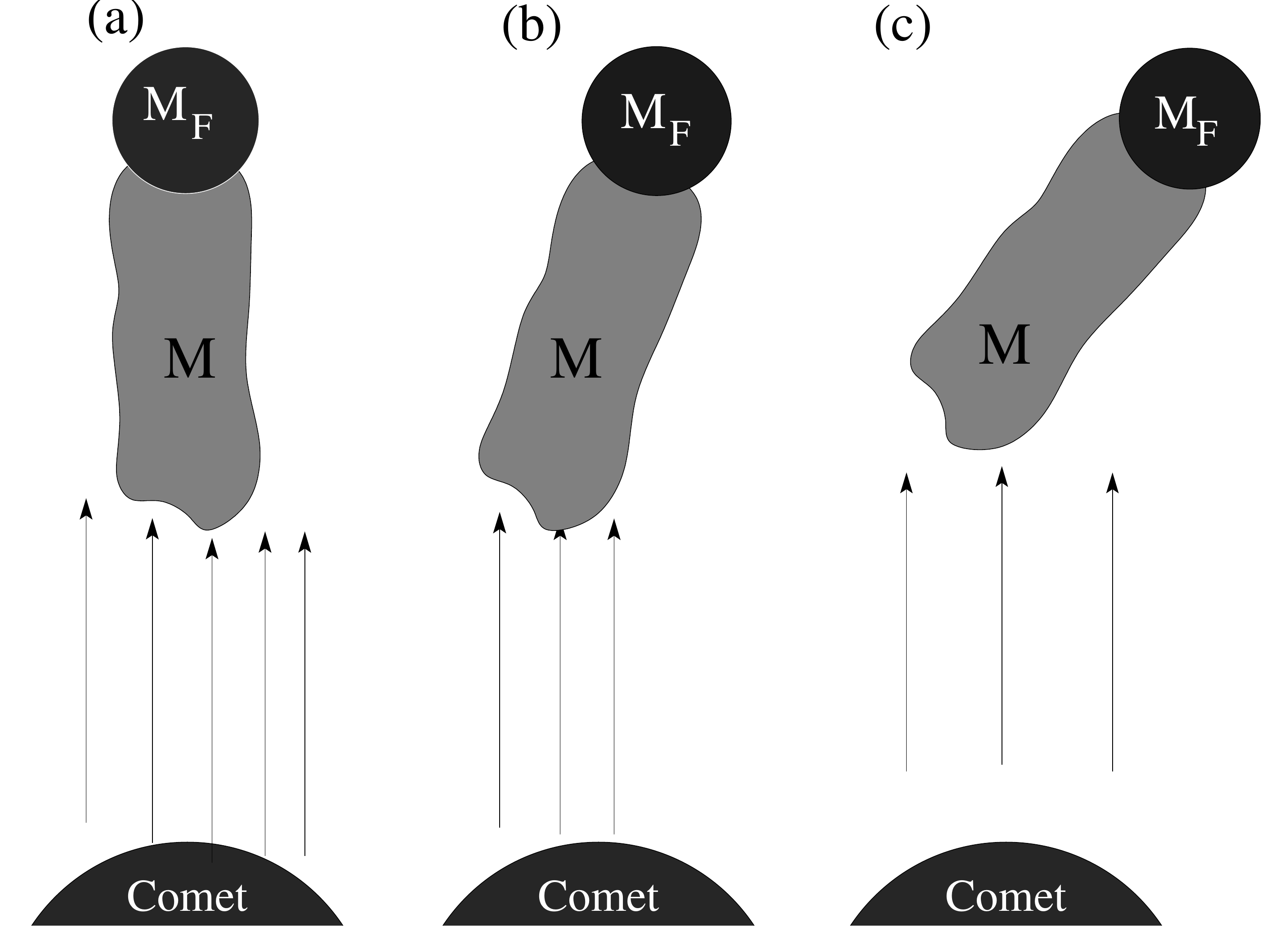}
\caption{The fractal aggregate breaking off from cometary fragment.  "M" denotes the fractal aggregate, while "M$_{\mbox{F}}$" denotes the cometary fragment. \label{breakoff}}. 
\end{figure}

\section{Ejection into interstellar space} 

Once formed, the CFA's very low density means that it is highly susceptible to radiation pressure.
The impact of radiation pressure is measured by the parameter $\beta$, defined as the ratio of the radiation 
pressure force to the gravitational force.  The radiation pressure at $R$ is given by  $P_{rad} = L_{Sun} / (4 \pi R^2 c)$,  where $L_{sun} = 4 \times 10^{26}$ W is the Sun's luminosity and $c = 3 \times 10^8$ m/s is the speed of light.  We calculate $\beta$ according to 
$\beta = \pi a b P_{rad} R^2 / \left(G M_{Sun} M_{CFA}\right) = 1.7 \times 10^{-3}$,
where $M_{Sun} = 2 \times 10^{30}$ kg is the solar mass, and $P_{rad} = 0.5~\mu$Pa at $R = 1$ AU.
Since long-period comets are already on parabolic (or nearly parabolic) orbits, their ejecta can be blown 
into unbound orbits by radiation pressure even when $\beta < 1$  (\cite{kresak1976}).  Observations have shown that comet fragments tended to lie in the anti-solar direction (\cite{desvoivres2000, ishiguro2009}), consistent with a sensitivity to radiation pressure.

If the CFA is released 
at heliocentric distance $R_{release}$, the criterion for unbound orbits is
$R_{release} \le 2~a_{s} \beta$,  where $a_s$ is the orbital semimajor axis (\cite{kresak1976, mukai1989}).
For $\beta = 10^{-3}$, $R_{release} = 1$ AU would require $a_{s} > 300$ AU for escape.  
Since this condition is satisfied by most long-period comets, the most likely fate for long-period CFAs is to escape to interstellar space. 

The few long-period CFAs not blown out by radiation pressure still have short lifetimes.  Inheriting their parent comets' orbits, 
they will make only a handful of returns to the inner solar system before being ejected into unbound orbits (\cite{wiegert1999}).  
We point out that very low activity long-period comets (the "Manx comets") have already been 
discovered (e. g., C2016 VZ18 and C2018 EF9) (\cite{molnar-bufanda2019}); further observations might reveal whether these comets might have a fractal nature. 

In theory, CFAs could also be produced by short-period comets; however, this would be rare due to the much 
reduced level of activity of these comets.  
Due to their many passages through the inner solar system, most short-period comets have built up insulating mantles
that choke off sublimation  (\cite{rickman1990}), resulting in a small fractional active surface area $f \le 0.2$  (\cite{fernandez2005}).  
For illustration, the mean free path for $1 \mu$m particles from a $f = 0.2$ comet is plotted in Figure 2, and the Figure
 does indeed show that the accretion condition is not satisfied.   If short-period CFAs exist,
like their parent comets, their lifetimes are also expected to be limited to $\sim 10^4$ years by planetary perturbations (\cite{disisto2009}). 
However, a CFA's low density makes it very vulnerable to the YORP torque, as we have already seen with 'Oumuamua (\cite{flekkoy19}).
Depending on the CFA's orbit and direction of its spin axis, its lifetime against rotational fission may well be 
shorter than its dynamical lifetime. 

\section{Statistics} 

Can interstellar CFAs explain 'Oumuamua's discovery statistics, namely 1 CFA  in $\sim 4$ years of PanSTARRS observations?
We estimate the number of 'Oumuamua-size CFAs as follows. 
Based on the 150+ fragments of the short-period
comet Schwassmann-Wachmann B3, a 1-km radius nucleus could produce $\sim 100$ 10-m size fragments (\cite{ishiguro2009}).
Assuming that the CFAs are distributed with an isotropic
velocity distribution throughout our galaxy, the rate at which CFAs are detected is then (\cite{jewitt2003})

\be
dN/dt = f_{grav}~n_{CFA}~\pi R_c^2~\Delta V. 
\label{dNdt}
\ee

\noindent Here $f_{grav}$ is the gravitational focusing factor, $n_{CFA}$ is the CFA number density, $\Delta V$ is the relative velocity,
and

\be
R_c = 5~\textrm{AU} \left(a / \textrm{km} \right)^{1/2}
\label{Rc}
\ee

\noindent is the limiting distance where an object with radius $a$ is detectable.  
The gravitational focusing factor is given by $f_{grav} = 1 + v_{esc}^2 / v_{\infty}^2$,  where $v_{esc}$ is the Sun's escape velocity, and $v_{\infty}$ is the velocity at infinity of the CFA.  Adopting the Sun's escape velocity at 1 AU, $v_{esc} = 42$ km/s, and 'Oumuamua's $v_{\infty} = 25$ km/s yields $f_{grav} \sim 4$.  
The fraction of Oort cloud comets with perihelion $q \le 3$ AU 
is $7\%$, with most of these comets 
making on average 6 returns before being lost to unknown physical ("fading") mechanisms  (\cite{wiegert1999}).
If we assume $N_s = 10^{11}$ stars in the galaxy, each of them having an Oort cloud with $N_O=10^{12}$ comets, 
the CFA number density is

\be
n_{CFA} = (0.07) ~  (6) ~ N_s N_O N_{CFA} /V_{gal},
\label{n_CFA}
\ee

\noindent   where $N_{CFA}$ is the number of CFAs per comet, and $V_{gal} = 4.5 \times 10^{27}$ AU$^3$ is the volume of the galaxy.  Equation \ref{n_CFA} can be rewritten as $n_{CFA} = 9 \times 10^{-4}~\left(N_{CFA} /100 \right)~\rm{AU}^{-3}$, 
and letting $a = 230$ m and $\Delta V$ = 25 km/s in Equation \ref{Rc}, the discovery rate is then 
$dN/dt \sim 0.35  \left(N_{CFA} /100 \right)$ year$^{-1}$.   If CFAs break up into secondary fractal bodies, the discovery rate may be even larger.  This is a crude order-of-magnitude calculation, but it seems that the propensity for comets to fragment might be capable of supporting 'Oumuamua's discovery statistics.

\section{Conclusion}

We present a model that shows how the interstellar object 'Oumuamua could have formed as a fractal aggregate ("dust bunny") in the inner coma of an exo-Oort cloud comet.  The formation of the fractal is made possible by two important factors: 1) the presence of cometary fragments as accretion sites, and 2) the fractal's ability to absorb impacts during the accretion process.  After forming, the fractal eventually separates from the fragment due to hydrodynamic stress, then is ejected into interstellar space by radiation pressure.  Our model reproduces the short dimension of 'Oumuamua, and the object's elongated shape (although the model cannot distinguish between a cigar or disk shape).  The model could also explain the discovery rate of 'Oumuamua, depending on the number of fragments produced per exo-Oort comet.  Finally, it predicts the existence of fractal bodies in the solar system, which may one day be identified.

\acknowledgments
We are grateful to the referee for helpful comments that improved the paper.  We also thank the Research Council of Norway through its Centres of Excellence funding scheme, project number 262644, and the LIA France-Norway D-FFRACT.

\bibliography{all}
\bibliographystyle{aasjournal}
\end{document}